\documentclass[aps,prb,preprint,showpacs,superscriptaddress]{revtex4-1}
\usepackage{graphicx}
\usepackage{epstopdf}
\usepackage{dcolumn}% Align table columns on decimal point
\usepackage{bm}% bold math
\usepackage{amsmath}
\usepackage{amsfonts}
\usepackage{amssymb}
\usepackage{natbib}
\usepackage[utf8]{inputenc}

\usepackage{ulem}
\begin{document}

%\preprint{}

%\title{Cloaking the focus of surface acoustic waves: swirling SAW synthesis by inverse filter} 
\title{Synthesis of anisotropic swirling surface acoustic waves by inverse filter, towards integrated generators of acoustical vortices} 

\author{Antoine Riaud}
\affiliation{Institut d'Electronique, de Micro\'electronique et Nanotechnologie (IEMN), LIA LICS, Universit\'{e} Lille 1 and EC Lille, UMR CNRS 8520, 59652 Villeneuve d'Ascq, France}
\affiliation{Sorbonne Universit\'{e}s, UPMC Univ Paris 06, CNRS, UMR 7588,  Institut des NanoSciences de Paris (INSP), F-75005, Paris, France}

\author{Jean-Louis Thomas}
\affiliation{Sorbonne Universit\'{e}s, UPMC Univ Paris 06, CNRS, UMR 7588,  Institut des NanoSciences de Paris (INSP), F-75005, Paris, France}

\author{Eric Charron}
\affiliation{Sorbonne Universit\'{e}s, UPMC Univ Paris 06, CNRS, UMR 7588,  Institut des NanoSciences de Paris (INSP), F-75005, Paris, France}

\author{Adrien Bussonnière}
\affiliation{Institut d'Electronique, de Micro\'electronique et Nanotechnologie (IEMN), LIA LICS, Universit\'{e} Lille 1 and EC Lille, UMR CNRS 8520, 59652 Villeneuve d'Ascq, France}

\author{Olivier Bou Matar}
\affiliation{Institut d'Electronique, de Micro\'electronique et Nanotechnologie (IEMN), LIA LICS, Universit\'{e} Lille 1 and EC Lille, UMR CNRS 8520, 59652 Villeneuve d'Ascq, France}

\author{Michael Baudoin}
\email{Corresponding author: michael.baudoin@univ-lille1.fr}
\homepage{http://films-lab.univ-lille1.fr}
\affiliation{Institut d'Electronique, de Micro\'electronique et Nanotechnologie (IEMN), LIA LICS, Universit\'{e} Lille 1 and EC Lille, UMR CNRS 8520, 59652 Villeneuve d'Ascq, France}

%\affiliation{UPMC Univ Paris 06, UMR 7588, Institut des NanoSciences de Paris (INSP), F-75005, Paris, France}

\date{\today}

\begin{abstract}
From radio-electronics signal analysis to biological samples actuation, surface acoustic waves (SAW) are involved in a multitude of modern devices. Despite this versatility, SAW transducers developed up to date only authorize the synthesis of the most simple standing or progressive waves such as plane and focused waves. In particular, acoustical integrated sources able to generate acoustical vortices (the analogue of optical vortices) are missing. In this work, we propose a flexible tool based on inverse filter technique and arrays of SAW transducers enabling the synthesis of prescribed complex wave patterns at the surface of anisotropic media. The potential of this setup is illustrated by the synthesis of a 2D analog of 3D acoustical vortices, namely "swirling surface acoustic waves". Similarly to their 3D counterpart, they appear as concentric structures of bright rings with a phase singularity in their center resulting in a central dark spot. Swirling SAW can be useful in fragile sensors whose neighborhood needs vigorous actuation, and may also serve as integrated transducers for acoustical vortices. Since these waves are essential to fine acoustical tweezing, swirling SAW may become the cornerstone of future micrometric devices for contactless manipulation .   
\end{abstract}

\pacs{}
%\keywords{}

\maketitle

%\section{Introduction\label{introduction}}

\section{Introduction \label{intro}}

Surface acoustic waves (SAW) have become the cornerstone of micro-electro-mechanical-systems (MEMS). SAW are not only useful in delay lines and convolution filters \cite{Royer_Dieulesaint2}, but can also monitor temperature variations, strain \cite{Tire_wireless_sensor}, magnetic fields \cite{Spin_acoustics1,Spin_acoustics2} and even chemical or biological composition \cite{SAW_sensor_review,AcousticWavesSensors}. More recently, the growing field of microfluidics has expressed tremendous interest towards SAWs \cite{DungReview,YeoFriendReview}, due to their versatility for droplet actuation \cite{franke2009surface,abc_wixforth_2004,sab_renaudin_2006,apl_baudoin_2012}, atomization \cite{jjap_chono_2004,pof_yeo_2008}, jetting \cite{jjap_shiokawa_1990,prl_tan_2009} or mixing  \cite{prl_frommelt_2008} but also bubbles, particles and cells manipulation and sorting \cite{BruusSettnes,Huang1, Thibault1, bussonniere2014cell}. 

\begin{figure}[htbp]
\includegraphics[width=80mm]{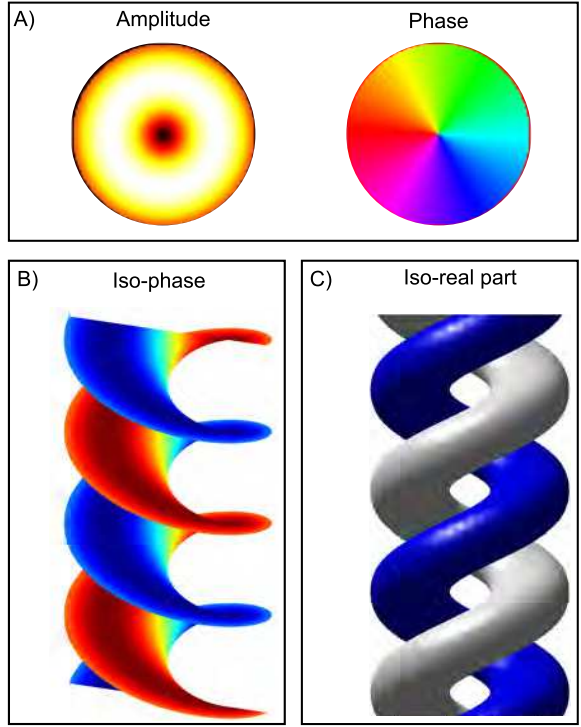}
\caption{A particular example of isotropic dark beams: the Bessel beams (eq. \ref{eq: iso_bessel_beam} in part \ref{seq: anisotropic_bessel}) with $l=1$, $k_z = 1$ and $k_r=1$. A) Beam cross-section with complex phase and amplitude. B) Iso-phase surfaces at $l\theta-k_z z = 0$ and $l\theta-k_z z = \pi$ in red and blue respectively. C) Iso-surface of $\Re (W_l) = -0.3$ and $+0.3$ in blue and white respectively.} \label{fig: IsotropicVortex} %In red and blue, the beam amplitude is traced over the isophases $0$ and $\pi$. In the inset are the isosurfaces  respectively.}  %A figure caption. The figure %captions are automatically numbered.}
\end{figure}

In order to improve their actuation power, researchers have dedicated years of efforts to SAW focusing, with significant success\cite{SAW_focusing1989,Laude2008,SAW_focusing2005,SAW_focusing2008}. An important side-product of this quest was the introduction of anisotropic zero-order Bessel functions \cite{Laude2006}. These functions are notorious in the fields of electronics \cite{Grillo2015,Electron_vortex1,Electron_vortex2}, optics \cite{Allen1992,He1995} and acoustics \cite{Durnin1987,jasa_hefner_1999,MarchianoThomas2005} for describing Bessel beams (fig. \ref{fig: IsotropicVortex}), a large family of vortical waves spinning around a phase singularity. In this singular point, destructive interferences lead to the total cancellation of the beam amplitude. The resulting dark core is one of the most attractive feature of Bessel beams as it can trap particles \cite{Anhaeuser2012,Marston2006,Riaud2014,Baresch2014}. In the present study, we aim at generating a two-dimensional version of acoustical vortices, called for convenience swirling surface acoustic waves.

%[Literature on plasmonic vortex:
% Surface plasmon polaritons generated by optical vortex beams
% Coupling of spin and angular momentum of light in plasmonic vortex, 
% Synthesis and Dynamic Switching of Surface Plasmon Vortices with Plasmonic Vortex Lens]

 In two dimensions, swirling SAW would appear as a dark spot circled by concentric bright rings of intense vibrations. It is tantamount to cloak the focus of surface acoustic waves, allowing vigorous actuation of the direct neighborhood of fragile sensors. Furthermore, SAW easily radiate from a piezoelectric solid to an adjacent liquid, simply by diving the transducer in the fluid. Swirling SAW could therefore serve as integrated acoustical tweezers \cite{Huang1,Thibault1}. This would solve one of the major shortcomings of advanced point-wise acoustical tweezers \cite{Drinkwater1,Drinkwater2,Baresch2014}, which are complex mechanical assemblies of numerous individual transducers, whereas the present swirling SAW generators are obtained by metal sputtering and photolithography on a single piezoelectric substrate. The radiation of swirling SAW in adjacent liquid might also be used to monitor cyclones-like flows \cite{Riaud2014} in cavities by using a nonlinear effect called acoustic streaming.

Two difficulties must be overcome prior to the generation of swirling SAWs. First, although SAW synthesis is well mastered for single transducers radiating in specific directions of piezoelectric materials, the design of interdigitated transducers arrays (IDTA) surrounding a control area is much more complex. Indeed, piezoelectricity happens at the expense of isotropy, as better piezoelectric coupling requires larger substrate anisotropy \cite{Royer_Dieulesaint}. Anisotropy considerably complicates the SAW propagation, leading to direction-dependent wave velocity, coupling coefficient and beam stirring angle (non-collinear wave and energy vectors). Thanks to recent mathematical developments \cite{GreenSAW1995a,GreenSAW1995b,GreenSAW1999,Laude2006}, SAW far field propagation is better handled nowadays. Nevertheless, these methods require an accurate depiction of the target field in order to design the generator. The second difficulty is to define exactly what a swirling SAW is, especially in an anisotropic medium. Since these waves are the fragile result of destructive interference, extreme care must be taken in computing their propagation.

In the present study, we use an adaptive field synthesis method in order to tackle the first issue.  For this purpose, a sample of piezoelectric material is covered with a circular array of 32 independent transducers actuated by a programmable electronic. Then, its vibrations are monitored by a Michelson interferometer. The exact input was computed by an advanced calibration procedure called inverse filter \cite{Tanter2000,MarchianoThomas2005,Baresch2014}.

Getting rid of the issue of emitter design, we efficiently focused on the definition of swirling surface acoustic waves. Our theoretical work was essentially guided by Laude et al. \cite{Laude2006}, who unveiled a zero-order anisotropic Bessel. In a different context (multipole expansion of electro-magnetic waves for numerical computation), Piller and Martin proposed a comprehensive extension of Bessel functions to anisotropic media \cite{Piller1998}. Our theoretical investigation, described in the first part of the paper, uses the concepts of slowness surface and angular spectrum to fill the gap between Piller and Martin mathematical expression and surface acoustic waves. The next part of the paper describes our experimental setup, from the transducers design to the SAW measurements. The third part explains how we designed the emitter signal in order to synthesize swirling SAW. It provides the key steps of the inverse filter method adapted to the propagation of surface acoustic waves. Finally, a fourth section exhibits some experimental swirling surface acoustic waves. To the best of our knowledge, it is the first time focused surface acoustic waves are synthesized with a phase singularity and the associated dark spot.

%Sensors for chemical and biological applications UPMC

%In the present paper, we borrow a mathematical function proposed earlier by Piller and Martin as an anisotropic generalization to Bessel functions, and extend it to surface acoustic waves. The complex problem of emitter design is considerably aleviated using 32 elementary programmable sources controlled by inverse filter. This experimental setup provides us a versatile laboratory to synthetize for the first time anisotropic dark surface acoustic waves.  

\section{Definition of an anisotropic Bessel function}\label{seq: anisotropic_bessel}

%Extension of the generalized multipole technique to anisotropic medias

%The definition of an anisotropic dark beam is not straightforward. In the special case of Bessel beams, the isotropic versions arise naturally in any situation involving a cylindrical geometry. Somehow, the present question is tantamount to defining what is a cylinder/a circle in an anisotropic medium. In any material, the dispersion relation is figured by a set of three curves: the slowness, wave and velocity contour.  

A classical solution to the wave equation in isotropic media is known as the Bessel beam:

\begin{equation}
W_l e^{-i \omega t} = J_l(k_r r)e^{il\theta + i k_z z -i \omega t} = W_l^0e^{ i k_z z -i \omega t}
\label{eq: iso_bessel_beam}
\end{equation}

In this equation, $r$, $\theta$, $z$, $t$, $l$, $W_l$, $J_l$, $k_r$, $k_z$, $\omega$, $W_l^0$ stand respectively for the axisymmetric coordinates, the time, the topological order, the complex wave field value, the $l^{\text{th}}$ order Bessel function, the radial and axial parts of wave vector, the angular frequency and the isotropic swirling surface acoustic wave complex value.

Most undergraduate courses introduce the Bessel function through series development. In contrast to more common waves,  surface acoustic waves are the result of linear algebraic-partial-differential equations \cite{Royer_Dieulesaint}, a class of problem where the use of series development, if not totally inefficient, would at best be cumbersome and inelegant.
%In the present study, surface acoustic waves are sometimes regarded as guided waves. They propagate unattenuated at the surface of solids, a condition which requires that no power is transmitted from the wave to the solid bulk. Since the normal velocity does not vanish at the interface, this condition can only be fulfilled if the longitudinal and transverse deformation can combine to balance the inertial stress. 
%light waves are guided waves on the existence cone

The slowness surface and angular spectrum \cite{Fagerholm1996} constitute the basic blocks for building an anisotropic Bessel beam. Hence, we will first briefly review these concepts, and then use them to design an anisotropic swirling SAW. The main idea of these tools is to reduce the problem to a superposition of plane waves.  For each single direction and frequency, we solve the 1-D propagation equation, which reduces the partial differential equation to a set of ordinary differential equations, whose integration is straightforward.  

In the following, we will work at a given frequency and omit the term $e^{-i\omega t}$ for clarity. In isotropic materials such as water, the wave speed of sound or light is independent of the direction of propagation. Consequently, the magnitude of the wave-vector $k = 2\pi/\lambda$ is also a constant, and its locus versus the direction of propagation is a sphere called the slowness surface. Conversely, in the case of an anisotropic material, the wave speed depends on the direction, and so is the wave-vector. In the reciprocal space of a 3D media, we call $\Phi$ the azimuth and $\kappa_z$ the altitude in cylindrical coordinates, so the wave-vector reads $\boldsymbol{k}(\phi,\kappa_z)$. The locus of this wave-vector, still called the slowness surface, then results in non-spherical shapes depending on the anisotropy of the material \cite{Royer_Dieulesaint}. Bessel beams propagate along a specific axis $\boldsymbol{z}$. Consequently, discussions on the surface slowness will often refer to $k_r(\phi,\kappa_z)$, the projection of $\mathbf{k}$ on the plane normal to the propagation axis. The axial and radial component of the wave-vector $k_z$ and $k_r$ are linked by the direction-wise dispersion relation:
\begin{equation}
	k_z(\phi,\kappa_z)^2 + k_r(\phi,\kappa_z)^2 = \frac{\omega^2}{c(\phi,\kappa_z)^2}
	\label{eq: disp_relation}
\end{equation} 
In equation (\ref{eq: disp_relation}), we wrote the coordinates in the reciprocal space $\kappa_i$ to distinguish them from $k_i$ which refer to the dispersion relation of the wave and are given physical quantities. For instance, $\kappa_z$ can take any value whereas $k_z$ is defined only in a closed interval ($k_z \in [-\omega/c , +\omega/c]$ for an isotropic medium). 

The angular spectrum is a multidimensional generalization of the Fourier transform. Since Fourier pioneering work, it is known that any field can be resolved into a sum of sinusoidal functions. The angular spectrum is a recursive application of the Fourier transform over all the dimensions of the media:
\begin{equation}
\begin{split}
f(x,y,z) = &\\
\int_{-\infty}^{+\infty} \int_{-\infty}^{+\infty} \int_{-\infty}^{+\infty} & F(\kappa_x,\kappa_y,\kappa_z) e^{i \kappa_x x}d\kappa_x e^{i\kappa_y y}d\kappa_y e^{i\kappa_z z}d\kappa_z
\end{split} 
\label{eq: Fourier cartesian}
\end{equation}

We can re-arrange the terms in the exponential in order to get $\exp(i(\kappa_x x + \kappa_y y + \kappa_z z))$ such that equation (\ref{eq: Fourier cartesian}) can be interpreted as a sum of plane waves. This means that any physical field in the media at a given frequency can be seen as a combination of plane waves and therefore must satisfy the dispersion relation, or equivalently lie on the slowness surface. In this regard, the slowness surface provides a frame for the wave landscape, and choosing the angular spectrum $F(\kappa_x,\kappa_y,\kappa_z)$ amounts to applying the color (complex phase and amplitude) on this frame.

If we express the previous angular spectrum not in Cartesian coordinates but in cylindrical ones, we get:
\begin{equation}
\begin{split}
f(r,\theta,z) = &\\
\int_{-\infty}^{+\infty}\int_{-\pi}^{+\pi}\int_{0}^{+\infty}& F(\kappa_r,\phi,\kappa_z) e^{i\kappa_r r \cos(\phi - \theta)}\kappa_r d\kappa_r d\phi e^{i\kappa_z z}d\kappa_z 
\end{split}
\label{eq: Fourier cylindrical}
\end{equation}

In this expression, the variables $\kappa_r,\phi,\kappa_z$ refer to the spectral domain whereas $r,\theta,z$ belong to the spatial one.  
In order to satisfy the dispersion relation, we know that $F$ must vanish anywhere except on the slowness surface, so  $F(\kappa_r,\phi,\kappa_z) = h(\phi,\kappa_z)\delta(\kappa_r-k_r(\phi,\kappa_z))$, with $k_r$ the magnitude of the wave-vector in the $(x,y)$ plane and $h$ an arbitrary function of $\phi$ and $\kappa_z$. This reduces the set of waves that can be created in the media:
\begin{equation}
\begin{split}
f(r,\theta,z) = &\\
\int_{-\infty}^{+\infty}\int_{-\pi}^{+\pi} & h(\phi,\kappa_z) e^{i k_r(\phi,\kappa_z) r \cos(\phi - \theta)}k_r(\phi,\kappa_z) d\phi e^{i\kappa_z z}d\kappa_z 
\end{split}
\label{eq: complicated_cyl_angul_spec}
\end{equation}

At a given $\kappa_z$, the integral in equation (\ref{eq: complicated_cyl_angul_spec}) is the product of two terms: the first one $ e^{i k_r(\phi,\kappa_z) r \cos(\phi - \theta)}$ can be reduced to a sum of plane waves thanks to Jacobi-Anger expansion, while the second one $h k_r$ provides the color of each of these planes waves. 

We construct anisotropic Bessel functions by splitting the wave angular spectrum in a $\kappa_z$-independent part and extracting its coefficients. Since $\phi$ is the azimuth, it is a periodic function and we can expand $h k_r$ in Fourier series: $h k_r = \sum_{-\infty}^{+\infty} a_l(\kappa_z) e^{i l \phi}$. We then get:
\begin{equation}
\begin{split}
f(r,\theta,z) = &\\
 \int_{-\infty}^{+\infty} \sum_{l = -\infty}^{+\infty} & a_l(\kappa_z) e^{i\kappa_z z}\int_{-\pi}^{+\pi}  e^{i l\phi +  i k_r(\phi,\kappa_z) r \cos(\phi - \theta)} d\phi d\kappa_z
\end{split}  
\end{equation}
As mentioned earlier, the integral can be interpreted as a sum over all the $\kappa_z$ of some elementary functions. In these functions, $\kappa_z$ appears as a parameter instead of a variable. 

In order to highlight what in this expansion may be reminiscent of a Bessel, we need to write the integral expression of the Bessel function:
\begin{equation}
J_l(x) = \frac{1}{2\pi}\int_{-\pi}^{+\pi}  e^{i l \eta - i x \sin(\eta)} d\eta 
\end{equation}
A trivial change of variable $\eta =   \phi - \theta - \pi/2$ yields:
\begin{equation}
J_l(x) = \frac{1}{2\pi i^l}\int_{-\pi}^{+\pi}  e^{i l (\phi-\theta) + i x \cos(\phi - \theta)} d\phi 
\label{eq: isoBessel integral}
\end{equation}
We combine equation (\ref{eq: iso_bessel_beam}) and (\ref{eq: isoBessel integral}) to get the isotropic swirling SAW:
\begin{equation}
W^0_l(r,\theta) = \frac{1}{2\pi i^l}\int_{-\pi}^{+\pi}  e^{i l \phi + i k_r r \cos(\phi - \theta)} d\phi 
\end{equation}

By analogy with the isotropic equation, we define an anisotropic swirling SAW with a given $\kappa_z = k_z$ as:
\begin{equation}
\mathcal{W}^0_l(r,\theta) = \frac{1}{2\pi i^l}\int_{-\pi}^{+\pi}  e^{i l \phi + i k_r(\phi,k_z) r \cos(\phi - \theta)} d\phi 
\label{eq: anisotropic beam}
\end{equation}

The beam in equation (\ref{eq: anisotropic beam}) corresponds to the beam introduced by Piller and Martin \cite{Piller1998} for solving anisotropic multipole scattering problem, which augurs such anisotropic Bessel beam might be extremely widespread in Nature. 

Interestingly, any wave in an anisotropic media can be written as a combination of anisotropic Bessel beams $\mathcal{W}_l = \mathcal{W}^0_l e^{i \kappa_z z}$:
\begin{equation}
f(r,\theta,z)= \int_{-\infty}^{+\infty} \sum_{l=-\infty}^{+\infty} a_l(\kappa_z) 2 \pi i^l \mathcal{W}^0_l(r,\theta,\kappa_z) e^{i \kappa_z z} d\kappa_z 
\end{equation}

In the rest of the paper, we will use the inverse filter to generate anisotropic swirling SAW $\mathcal{W}^0_l$ on the surface of an anisotropic piezoelectric crystal.

\section{Experimental setup}

\begin{figure}[htbp]
\includegraphics[width=80mm]{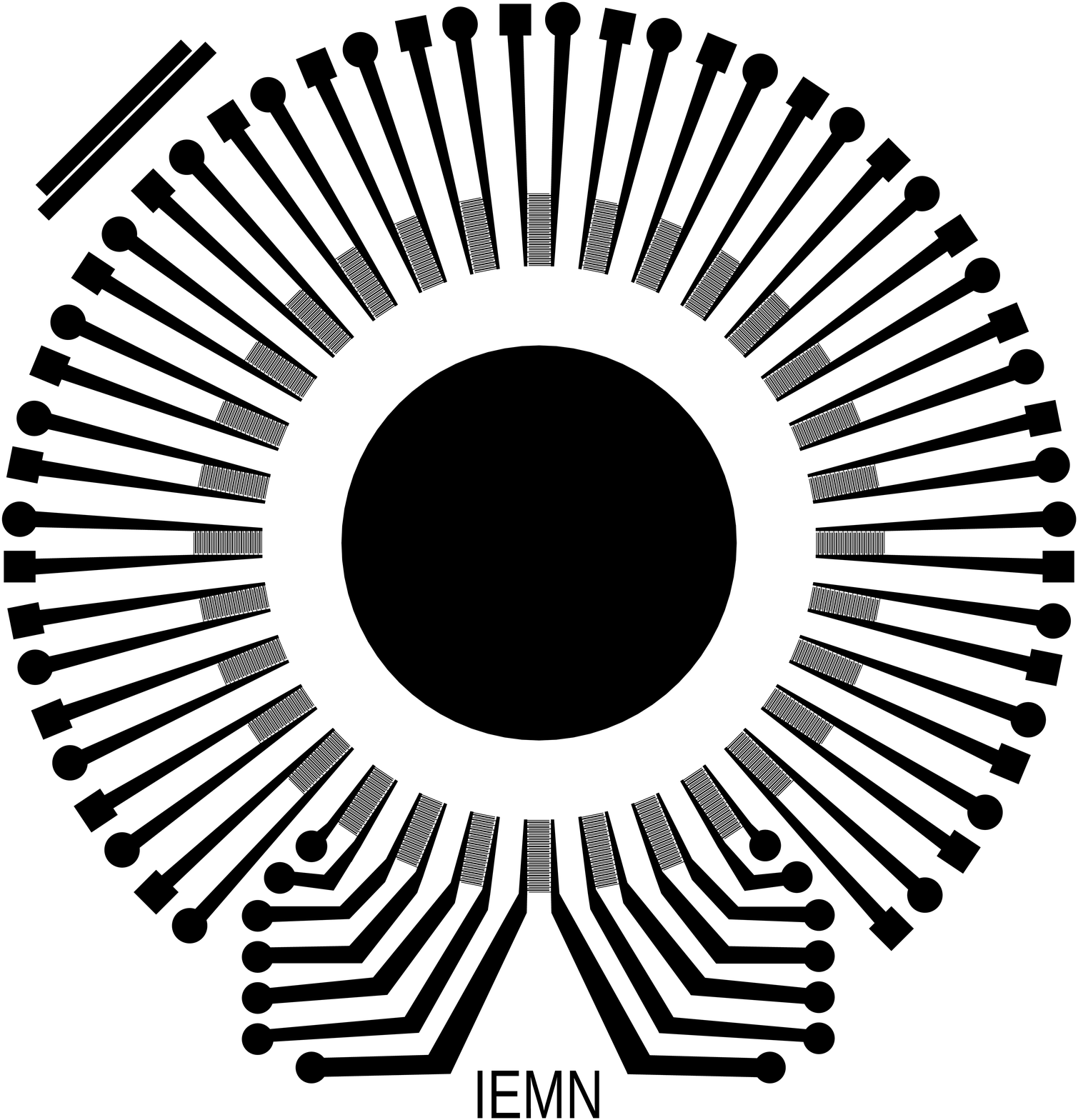}
\caption{Intedigitated transducer array used for generating the surface acoustic waves. The central black disc (25 mm diameter) is a gold layer acting as a mirror for interferometric measurements and materializes the maximum extent of the acoustic scene. Vector format image (available online) allows to visualize the fine structure of the electrodes.} 
\label{fig: IDT_array}  %A figure caption. The figure %captions are automatically numbered.}
\end{figure}

The experimental setup is designed to be as versatile as possible, in order to allow generating a wide variety of waves on an area called the acoustic scene. Starting from an X-cut of lithium niobate crystal, 32 unidirectional interdigitated transducers (SPUDT - IDT) were deposited on its periphery (see figure \ref{fig: IDT_array}). In order to widen the range of possible acoustic fields, every spot on the scene should be illuminated by all the transducers. This spatial coverage should be as uniform as possible on the acoustic scene. Furthermore, since any wave can be described as a combination of plane waves, it is essential to generate waves from a wide span of directions. Hence, the quality of the wave field synthesis critically depends on the span of plane waves provided by the source array in terms of incident angle, which is the angular spectrum coverage. The best way to achieve such optimal coverage is therefore to gather many sources from all the directions, and dispose them radially around a target spot which will be the acoustic scene. This spot should also be far enough away such that diffraction ensures a uniform spatial coverage. These notions of optimal coverage are detailed further in the next section and in the appendix \ref{sec: Appendix}. 

\begin{figure}[htbp]
\includegraphics[width=80mm]{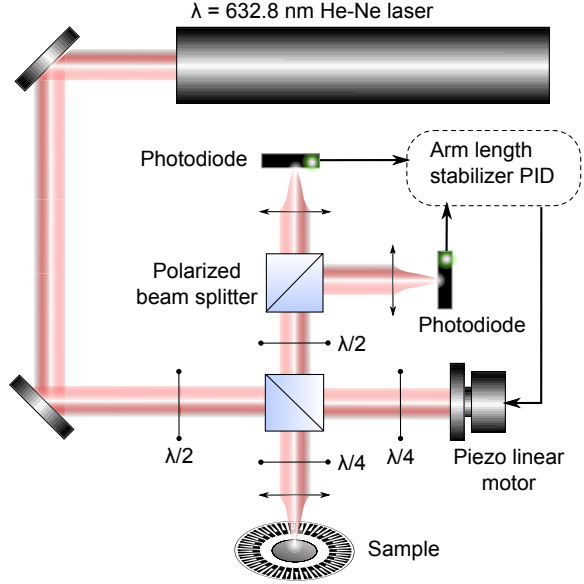}
\caption{Polarized Michelson interferometer used for scanning the displacement field associated with surface acoustic waves.} \label{fig: setup}  %A figure caption. The figure %captions are automatically numbered.}
\end{figure}

In order to measure the wave field on the acoustic scene, we placed the sample under the motorized arm of a polarized Michelson interferometer (Fig. \ref{fig: setup}). The poor reflection coefficient of lithium niobate was significantly increased by the deposition of a thin layer of gold on the acoustic scene ($\sim$ 200 nm).

\begin{figure}[htbp]
\includegraphics[width=80mm]{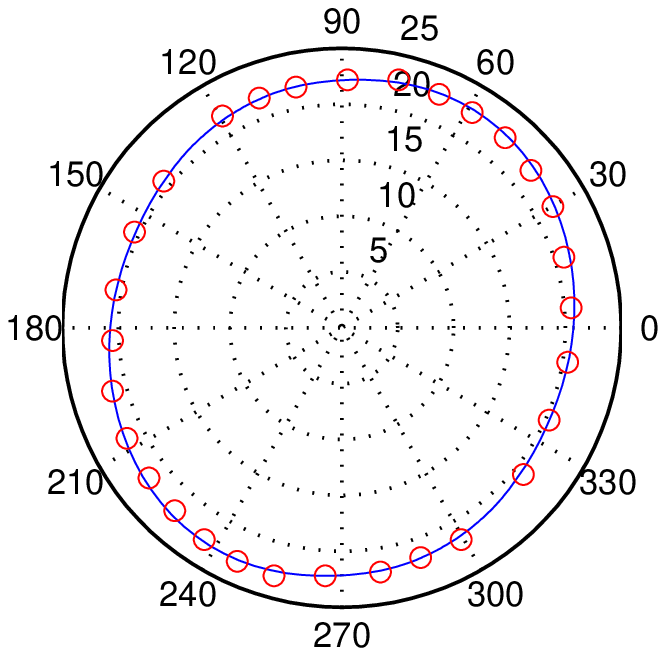}
\caption{Theoretical slowness contour (rad./mm) under a very thin gold layer \cite{CampbellJones} versus experimentally measured one.} 
\label{fig: experimental_slowness}  %A figure caption. The figure %captions are automatically numbered.}
\end{figure}

During the design of the IDTs, special care was given to the anisotropy of the lithium niobate substrate. Indeed, IDTs are high quality spatio-temporal resonant elements with a spatial period equal to the wavelength. Any deviation from the narrow resonant bandwidth results in very significant loss of efficiency \cite{Royer_Dieulesaint2}. We trace in figure \ref{fig: experimental_slowness} the slowness contour of lithium niobate measured on the gold layer at the working frequency of 12 MHz, and compare it to theoretical predictions \cite{CampbellJones}.

The knowledge of the dispersion relation provides the wave field radiated by a single point source \cite{Laude2006,Laude2008}:

\begin{equation}
G(r,\theta)\simeq A a(\bar{\phi})\frac{\exp\left(-i\omega r h(\bar{\phi}) - i \frac{\pi}{4}\textit{sign}(h''(\bar{\phi}))\right)}{\sqrt{\omega r \left|h''(\bar{\phi})\right|}}	
\label{eq: Green_Laude}
\end{equation}

With $a(\phi)$ the coupling coefficient between the field to measure and the electrical potential (obtained when solving the SAW equations \cite{CampbellJones,BouMatar2013}), $\bar{\phi}(\phi)$ the beam stirring angle, $h(\phi) = \cos(\phi-\theta)k(\phi)/\omega(\phi)$, $h'' = \frac{d^2h(\bar{\phi}(\phi))}{d \phi^2}$ related to the focusing factor. The beam stirring angle is solution of $h'(\bar{\phi}) =  0$. 

\begin{figure}[htbp]
\includegraphics[width=80mm]{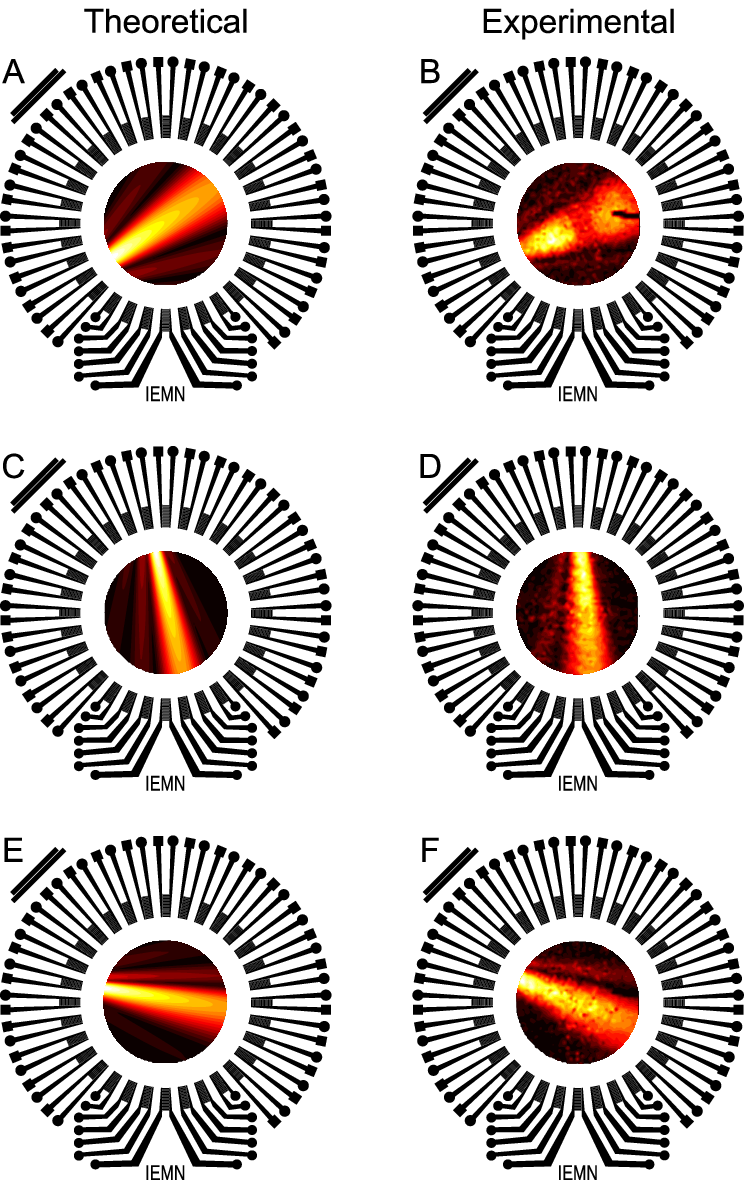}
\caption{Influence of anisotropy on the propagation of SAW generated by single electrodes. A,C,E) theoretical predictions, B,D,F) experimental measurements. A,B) beam widening, C,D) beam focusing, E,F) beam stirring. Color (available online) represents the beam relative intensity over the substrate, and is not indicative of the ratio of intensity between two different transducers.} 
\label{fig: diffraction_saw}  %A figure caption. The figure %captions are automatically numbered.}
\end{figure}

Thanks to superposition principle, we can use the Green function in eq. (\ref{eq: Green_Laude}) to compute the acoustic field radiated by our emitter arrays. The predictions are compared to experiments in figure \ref{fig: diffraction_saw}. Anisotropy strongly affects the SAW propagation, as we can observe beam widening (A,B), focusing (C,D) and stirring (E,F) depending on the beam direction.  

A wide-band high-power multichannel FPGA (Lecoeur Electronics) powers the 32 emitters with tailored numerical input. The input is specific to each desired wave-field and designed through the inverse filter method.

\section{Inverse filter theory} \label{seq: inverse_filter}

Inverse filter \cite{Tanter2000} is a very general technique for analyzing or synthesizing complex signals that propagate through arbitrary linear media. This method is especially suited for prototyping, because given a set of independent programmable sources, it finds the optimal input signal to get a target wave field. When used for this purpose, it is similar to computer generated holography in optics \cite{Heckenberg1992}. This method was previously shown to be among the most accurate ones for generating acoustical vortices in isotropic media \cite{MarchianoThomas2005}. 

%\begin{widetext}
\begin{figure*}[htbp]
\includegraphics[width=160mm]{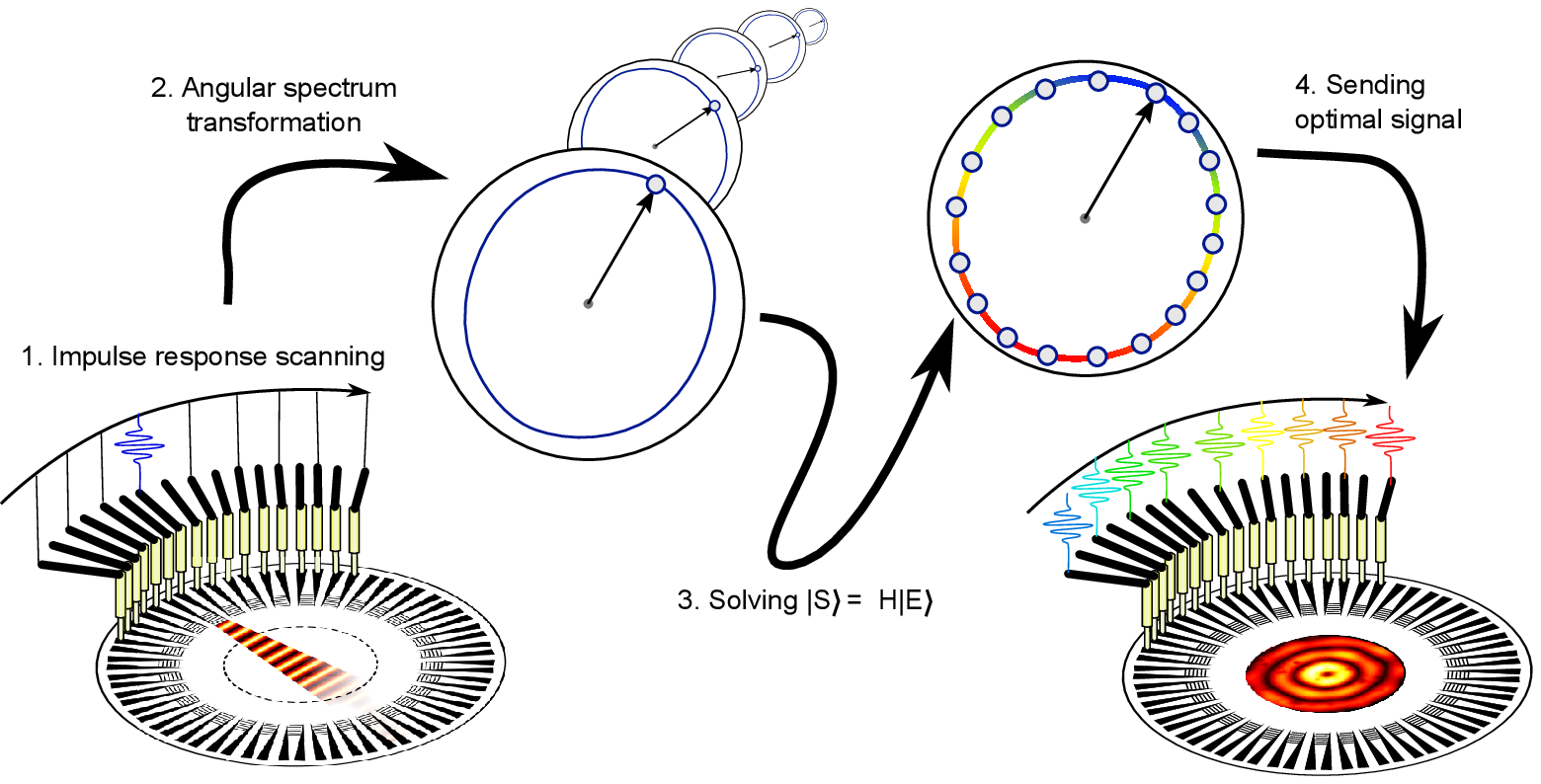}
\caption{Inverse filter flowchart. Inverse filter happens in four steps: (1) recording of the spatial impulse response (H matrix) for all transducers. (2) Transformation of the H matrix from spatial to spectral domain, where the response is sharper. (3) Computation of the optimal input $|\boldsymbol{E}\rangle $ for a desired output $|\boldsymbol{S}\rangle $ by pseudo-inversion of the matrix $\boldsymbol{H}$. (4) Generation of the signal from optimal input $|\boldsymbol{E}\rangle $.} \label{fig: inverse filter flowchart}  %A figure caption. The figure %captions are automatically numbered.}
\end{figure*}
%\end{widetext}

The method happens in four distinct stages (see figure \ref{fig: inverse filter flowchart}), (1) calibration of the transducers, (2 and 3) computation of the optimal input, and (4) actuation of the sound sources according to the optimal input.

In the current system, we used a set of 32 emitters, and an arbitrary number of control points evenly distributed on the acoustic scene. Their density is governed by Shanon principle: the distance between two points should not exceed $\lambda/2$. In our aquisition, we used a step of $\lambda/6 = 30$ $\mu$m. Moving the arm of the interferometer, we were able to reach individually each of these measurement points. If we call $e_i$ the temporal input of emitter $i$, and $s_j$ the temporal output of control point $j$ located on $\{x_j,y_j\}$, we have for any linear media:
\begin{equation}
s_j =\sum_{i} h_{ij}*e_i
\end{equation}
Where $*$ refers to the convolution product, and $h_{ij}$ is the time-response at control point $j$ to an impulse input at emitter $e_i$. In the spectral domain, $H_{ij} = \mathcal{F}(h_{ij})$ is the Fourier transform of the transfer function at control point $j$ of emitter $i$, and includes the propagation of the wave in the media. Using the matrix formalism, things get even simpler:
\begin{equation}
|\boldsymbol{S}\rangle = \boldsymbol{H} | \boldsymbol{E} \rangle
\end{equation}
In case where the transfer matrix is square and well conditioned, it can be inverted to determine the optimal input $| \boldsymbol{E} \rangle$ from a desired output $|\boldsymbol{S}\rangle$. However, the number of independent sources and control points is not necessarily the same, so $\boldsymbol{H}$ is generally not square and often ill-conditioned. In the appendix, we explain the reasons for this ill-conditioning from the perspective of the angular spectrum, and provide guidelines to minimize it. 

%previous versions of the inverse filter acted similarly to Point-source holograms, whereas Fourier-transform method is more efficient

As soon as the value of $|\boldsymbol{E}\rangle$ has been computed, the time-dependent input is obtained by inverse Fourier transform and sent to the FPGA amplifier to generate the wave field.

\section{Experimental results}

Bessel beams draw large interest for three main reasons: they do not diffract \cite{Durnin1987}, they carry a pseudo-orbital momentum {\cite{Allen1992,MarchianoThomas2003,Volke2008} and they exhibit a dark core (for nonzero order) \cite{Baresch2014,Drinkwater1}. In addition to these reasons, the zero order Bessel beam is the optimal beam focusing for a given aperture \cite{Laude2008}. In the following section, we start synthesizing a focused surface acoustic wave $\mathcal{W}_0^0$, and then some simple  first order swirling SAW $\mathcal{W}_1^0$. We will seize this opportunity to show the phase singularity and the associated dark spot. The size of the dark spot can easily be tuned, simply by changing the topological charge $l$, which is done in the third example with $7^{th}$ order swirling SAWs.

The zero order focused $\mathcal{W}_0^0$ Bessel wave phase and amplitude are traced on figure \ref{fig: W00}. It appears that theoretical fields and experimental ones are quite similar. In practice, we had to restrain the voltage amplitude of our instrument to about 10\% is order to get a linear response of the interferometer (upper bond is about 40 nm). For high actuation power, we estimated the displacement amplitude based on the second bright ring. When setting the voltage to about 50\%, we achieved displacement amplitude of nearly 180 nm.
%easier to explain than nonlinear analysis
%DL sur sin(2*pi*2*u/lambda) -> (4*pi*u/lambda)^2 << 6)
%from the second ring, displacement amplitude at the center is about 70 nm.
%repeat this experiment with lower power

\begin{figure}[htbp]
\includegraphics[width=80mm]{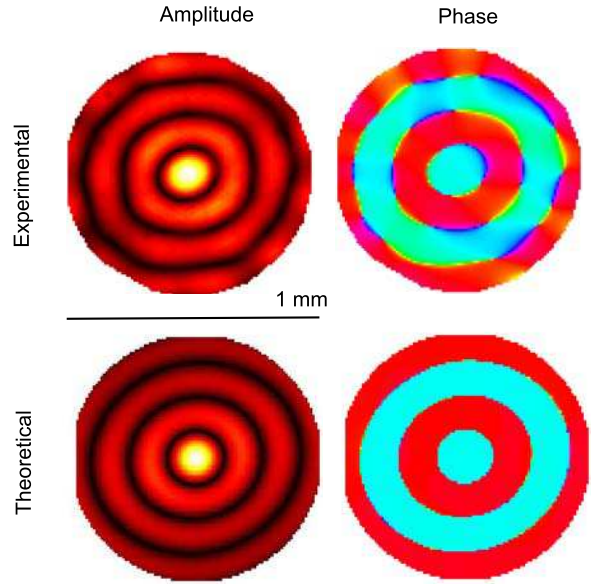}
\caption{Experimental and theoretically predicted zero order focused $\mathcal{W}_0^0$ Bessel wave phase and amplitude. Maximum experimental displacement is 40 nm.} \label{fig: W00}  %A figure caption. The figure %captions are automatically numbered.}
\end{figure}

Figure \ref{fig: W01} represents the first order dark beam $\mathcal{W}_1^0$ phase and amplitude. A dark core of zero amplitude with a diameter of 50 $\mu$m is clearly visible at the center of the vortex, and matches with a phase singularity. This area is contrasted by very bright concentric rings. Despite some blur in the experimental measurements, a good matching between theoretical and experimental vortices is achieved both on shape and phase.

\begin{figure}[htbp]
\includegraphics[width=80mm]{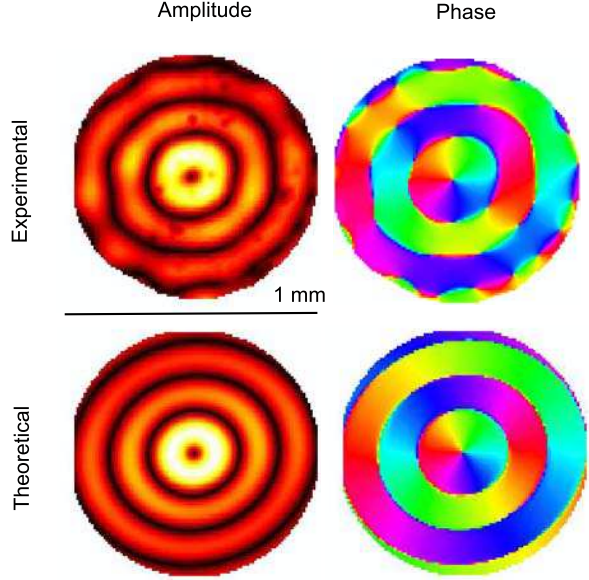}
\caption{Experimental and theoretically predicted first order  $\mathcal{W}_1^0$ Bessel wave phase and amplitude. Maximum experimental displacement is 36 nm.} \label{fig: W01}  %A figure caption. The figure %captions are automatically numbered.}
\end{figure}

Swirling SAW might be useful as integrated transducers for acoustical tweezers or micro-pumps. Tuning the topological order is essential to these applications for two reasons: it enlarges the first bright ring of the vortex (Olver formula \cite{Olver1951} in eq. \ref{eq: Olver}) and it increases the pseudo-angular momentum of the wave \cite{MarchianoThomas2003}. The second effect itself generates acoustic streaming with an azimutal flow velocity proportional to the topological charge \cite{Riaud2014,Volke2008}. 
\begin{equation}
j'_{l,1} = l(1+0.809\times l^{-2/3}) + O(l^{-7/3})
\label{eq: Olver}
\end{equation}

In this last example, we suggest a way to increase the ring radius while maintaining zero azimutal streaming and keep working at the resonance frequency of the electrodes. When two isotropic vortices of opposite charge are combined, they result in a circular stationary wave pattern. In the present case, we summed two seventh order contra-rotating acoustical vortices $\mathcal{W}_7^0+\mathcal{W}_{-7}^0$. The resulting field, shown in figure \ref{fig: W07f} exhibits a dark core with a diameter of about $500$ $\mu$m circled by a crown made of fourteen extrema of amplitude.

%In analogy to helicopters, we can combine two vortices of opposed topological order. 

%Here, we used a charge $l$ results in a near ten-fold increase of the dark core radius. This provides a practical mean to trap larger particles while working at a given frequency to ensure high efficiency of the interdigitated transducers. However, increasing vortex charge results in larger acoustic streaming \cite{Riaud2014}, unless in analogy to helicopters, two contra-rotative vortices are used as here.  

\begin{figure}[htbp]
\includegraphics[width=80mm]{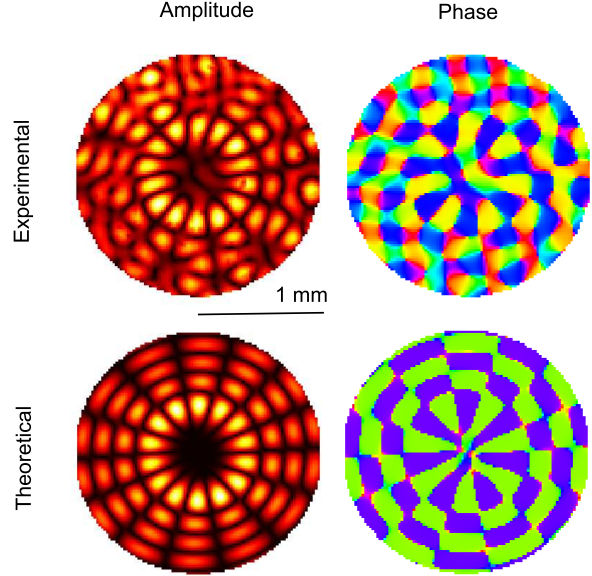}
\caption{Experimental and theoretical predictions of the combination of two seventh order vortices $\mathcal{W}_{\pm 7}^0$ of opposite charge. Maximum experimental displacement is 25 nm.} \label{fig: W07f}  %A figure caption. The figure %captions are automatically numbered.}
\end{figure}

\section{Conclusion}
%[reiterate the action, andswer the questions, demonstrate how it contributes to the larger problem]
In this report, we proposed an anisotropic SAW version of acoustical vortices, labeled swirling surface acoustic waves. This implied solving two difficulties: first, the generator had to be designed to accommodate anisotropic propagation, and second, we needed to define accurately what are swirling SAW. The first problem was alleviated using a programmable array of transducers controlled by a 2-dimensional version of the inverse filter, while the solution of the second problem confirmed earlier theoretical predictions. We synthesized swirling SAW of different topological charges and large magnitude of displacement. This successful generation provides a pathway for  integrated acoustical vortex generators on anisotropic substrates. Furthermore, since these beams are expected to radiate in any adjacent fluid, photolithography fabricated swirling SAW transducers may offer a credible alternative to current complicated acoustical tweezer devices made of mechanical assemblies of individual transducers. Beyond the specificities of acoustics, Bessel functions are very widespread in Nature and anisotropic Bessels may offer new analytical solutions for a broad class of linear anisotropic problems.   

%Geometrical description of the problem

%\begin{figure}[htbp]
%\includegraphics[width=80mm]{simulation_geometry_Renderer_painters.eps}
%\caption{Geometry of the simulation \textit{PML will be added, and domains will be numbered}} \label{fig: geometry}  %A figure caption. The figure %captions are
%automatically numbered.}
%\end{figure}

\appendix*

\section{Optimal conditioning of inverse filtering} \label{sec: Appendix}
Inverse filtering is a very versatile method to synthesize an optimal target field from a given number of transducers. 
As stated in section \ref{seq: inverse_filter}, the method is not exempt of poor-conditioning which would result in large errors in the synthesized field, but some guidelines can significantly improve the quality of field synthesis. 
The poor conditioning of inverse filter has two roots: (1) spectral outliers and (2) redundant sources. 

\subsection{Spectral outliers}

At a given frequency, the wave field must fulfill the dispersion relation, which is to have its angular spectrum lying on its slowness surface. When the acoustic scene is a surface (in 2D) or a volume (in 3D), this condition exactly happens. However, experimentally, there is always some noise introduced in the impulse response matrix, making it full rank (any spatial frequency can be created provided there is enough input power). Consequently, a first regularization is to remove the spectral outliers by sampling the target field not on a spatial manifold but on a spectral one, and along the slowness surface.

Nevertheless, if the acoustic scene is a line (in 2D) or a surface (in 3D), the spectral condition is relaxed. Indeed, the angular spectrum of the target field is only partially known due to the projection of the field along the line or the surface. In 3D for instance, if the synthesis happens on an \{x,y\} plane, the system knows the values of $k_x$ and $k_y$ but ignores the ones of $k_z$ which can then be freely chosen as long as the dispersion relation is fulfilled. In an isotropic medium, this results in $k_z = \pm \sqrt{\omega^2/c^2 - {k_x}^2-{k_y}^2}$. Note that in any case, $\omega^2/c^2 >{k_x}^2 + {k_y}^2$ which is the diffraction limit. Hence, spectral outliers in this synthesis appear beyond the $\lambda/2$ boundary.

A third example of spectral outliers is provided by piezoelectric generation on monocrystals. These substrates often exhibit a direction where the piezoelectric coupling coefficient sharply drops to zero. When this happens, no acoustic waves can be generated from this orientation and the associated angular spectrum coverage is barely zero. Once again, sampling the signal in the spectral space and excluding the zero-coupling directions avoids these outliers and allows an accurate synthesis. 

\subsection{Redundant sources}

In practice, many transducers are used to ensure an efficient spectral coverage. Above this threshold, adding even more actuators may result in poorer synthesis quality \cite{Tanter2000}. Indeed, from the inverse filter perspective, sound sources act like a family of vectors to combine in order to build a target field. When two sources are redundant, the inversion operator can take any linear combination of them, and this indetermination will be solved by comparing the measurement noise associated with each source. A smart way is therefore to regularize the reduced impulse response matrix obtained after removing the spectral outliers. The regularization can be achieved by a singular value decomposition. If two transducers are redundant, they will split in a singular value very close to zero and another one much more regular. Knowing the signal to noise ratio, it is then possible to discriminate which singular values originate from noise and which one are not \cite{Tanter2000}.

\bibliography{biblioa,AnisotropicVortexBibliography}

\begin{acknowledgments}
The authors would like to thank Silbe Majrab who developed and constructed the piezo-regulator of the interferometer, and Rémi Marchal who shared with us precious advices on the art of Michelson interferometry. This work was supported by ANR project ANR-12-BS09-0021-01.

\end{acknowledgments}

\end{document}